\documentclass[aps,pra,twocolumn,showpacs,amsmath,amssymb,preprintnumbers,superscriptaddress,10pt]{revtex4-2}

\usepackage{amsmath}
\usepackage{braket}
\usepackage{graphicx}
\graphicspath{{figures/}}
\usepackage{SIunits}
\usepackage{hyphenat}
\usepackage[bookmarks=false]{hyperref}
\usepackage{color}
\usepackage[capitalise]{cleveref}
\crefname{section}{Sec.}{Secs.}
\Crefname{section}{Section}{Sections}

\definecolor{pink}{RGB}{255,0,255}

\definecolor{red}{rgb}{0,0,1}

\begin{document}

\title{Nonclassicality of a delayed remote-controlled quantum computing model}

\author{Peicun Lin}
\affiliation{\mbox{College of Computer Science and Technology, National University of Defense Technology, Changsha 410073, China}}

\author{Dongyang Wang}
\affiliation{\mbox{College of Computer Science and Technology, National University of Defense Technology, Changsha 410073, China}}

\author{Yong Liu}
\affiliation{\mbox{College of Computer Science and Technology, National University of Defense Technology, Changsha 410073, China}}

\author{Anqi Huang}
\email{angelhuang.hn@gmail.com}
\affiliation{\mbox{College of Electronic Science and Technology, National University of Defense Technology, Changsha 410073, China}}
\affiliation{\mbox{College of Computer Science and Technology, National University of Defense Technology, Changsha 410073, China}}

\date{\today}

\begin{abstract}
Delegated quantum computing is likely to become the primary means for most people to access quantum computers in the future. However, these hardware inevitably operate beyond clients' control, raising concerns about potentially untrusted servers. A fundamental question thus arises---how can clients verify that the server is genuinely performing quantum computations? Here, we demonstrate that a class of remote-controlled quantum computing (RCQC) models presents the nonclassical behavior verified in a semi-device-independent way. To achieve this, with slight modifications, these models can be described by the prepare-and-measure scenario. By verifying the violations of dimension witnesses, classical causal models can be ruled out, thereby showing nonclassicality of the RCQC model. Remarkably, in the prepare-and-measure scenario, this class of RCQC models happens to exhibit reversed temporal order in quantum information processing. We also explicitly confirm the nonclassical behaviors of a specific 1-U-M RCQC model belonging to this class as an example. This work bridges the fundamental quantum theory with the practical task of quantum computing.
\end{abstract}

\maketitle

\section{Introduction}
\label{sec:intro}

Quantum computing has the potential capability to effectively solve computational problems that are currently intractable on classical computers, such as large integer factorization~\cite{Shor1995PolynomialTimeAF}, unsorted database search~\cite{Grover_1997}, and machine learning~\cite{QML}. This advantage in computational power is derived from the quantum mechanism---quantum state superposition, quantum entanglement, and so on. However, since quantum computers operate in highly specialized environments, it is both practically and economically motivated that the first generation of quantum computers will most likely act as servers, remotely providing quantum computing resources to users. Over the past decade, services have been provided for remote access to small-scale quantum processors over the Internet~\cite{Bristol,ESSDECR_2016}. Moreover, scholars have already utilized these platforms to conduct experiments~\cite{IBM_Exp1,IBM_Exp2,IBM_Exp3,IBM_Exp4}. These facts have borne out the speculation that the field of quantum computing will follow a path similar to classical cloud computing, namely quantum cloud computing~\cite{nguyen2024quantumcloudcomputingreview}. 

Since the cloud hardware is beyond clients' control, the quantum computing server may be untrusted. Protocols and models such as blind quantum computing~\cite{Broadbent_2009,Aharonov2010,reichardt2012classicalleashquantumsystem,Morimae_2013,Reichardt:2013mdb,HM15,Broadbent_2018,FHM18}, quantum homomorphic encryption~\cite{BJ15,DSS16,Qin2025}, and remote-controlled quantum computing (RCQC)~\cite{Qiang_2017,Wang:20} have been proposed to achieve secure delegated quantum computing and protect clients' privacy. Blind quantum computing protocols~\cite{Broadbent_2009,Aharonov2010,reichardt2012classicalleashquantumsystem,Morimae_2013,Reichardt:2013mdb,HM15,Broadbent_2018,FHM18} integrate notions of quantum computation with quantum cryptography to keep the structure of the computation delegated to the server being hidden. Quantum homomorphic encryption~\cite{BJ15,DSS16,Qin2025} allows quantum computations to be performed directly on encrypted data without prior decryption. In addition to these protocols, the RCQC models proposed in Refs.~\cite{Qiang_2017,Wang:20} achieve secure remote quantum information processing by concealing information about the unitary operator applied to the data state. Moreover, some of these protocols~\cite{Aharonov2010,reichardt2012classicalleashquantumsystem,HM15,Broadbent_2018,FHM18,Qin2025} provide mechanisms to verify that the untrusted server has performed computations as instructed by the client.

However, there is a lack of research on allowing the client to verify whether the server is truly performing quantum computation rather than returning an unreliable result with insufficient classical computing~\cite{reichardt2012classicalleashquantumsystem}. This is also an important assumption universally relied upon by the aforementioned protocols. Since the establishment of quantum mechanics, many studies have been conducted to verify the contradictions between classical hidden variable theories and quantum mechanics. Some renowned experiments among them are Wheeler's delayed-choice experiment~\cite{WHEELER19789,doi:10.1126/science.1136303} and its quantum variants~\cite{PhysRevLett.107.230406,doi:10.1126/science.1226755,Ionicioiu2018,Xiao-song2019}, in which the wave-like or particle-like behavior of a photon is decided after it has already entered the Mach-Zehnder interferometer. However, using tools from the field of causal inference~\cite{J.Pearl2009}, these counterintuitive experiments~\cite{WHEELER19789,doi:10.1126/science.1136303,PhysRevLett.107.230406,doi:10.1126/science.1226755} have been demonstrated to be classically explicable~\cite{PhysRevLett.120.190401}. Remarkably, based on the violation of the dimension witness~\cite{PhysRevLett.112.140407,PhysRevLett.105.230501}, Ref.~\cite{PhysRevLett.120.190401} devised a device-independent (DI) proof of nonclassical behaviors in a modified delayed choice experiment, which has been demonstrated and verified~\cite{PhysRevA.100.022111,PhysRevA.100.012115,PhysRevA.100.012114}. Could delegated quantum computing likewise exhibit nonclassical behaviors that admit DI proof?

Our main focus is to reply this question, which gives a positive answer. We demonstrate that a class of RCQC models proposed by Ref.~\cite{Qiang_2017,Wang:20} can be described by a prepare-and-measure (PAM) scenario, within which the nonclassicality of the model can be verified. In the PAM scenario, the RCQC model happens to exhibit reversed temporal order in quantum information processing---contrary to the traditional client-server architecture, the server can complete the computation first, after which the client determines the task delegated to the server. This feature is analogous to delayed-choice experiments. To emphasize it, we refer to it as the delayed RCQC model. Most importantly, we show that within the PAM scenario, nonclassical behaviors of the delayed 1-U-M RCQC model can be tested in a semi-device-independent (SDI) way. Specifically, given any model belonging to this class, we can test---similar to Ref.~\cite{PhysRevLett.120.190401}---the violations of dimension witnesses to rule out classical causal models based on some assumptions, thereby verifying that the model exhibits nonclassical behaviors. Moreover, we explicitly show nonclassical behaviors of a specific 1-U-M RCQC model proposed in Ref.~\cite{Wang:20}.

The structure of this paper is as follows. In \cref{sec:the DRC model and its nonclassicality}, we briefly introduce the RCQC model, and particularly focus on the 1-U-M RCQC model that can be described by a PAM scenario, and demonstrate how dimension witnesses for the PAM scenario can be used to verify whether this model exhibits nonclassical behaviors. In \cref{sec:a specific implementation of model}, we present a specific implementation of the delayed 1-U-M RCQC model and demonstrate its nonclassical behaviors. In \cref{sec:conclusion}, we summarize our work, outline potential applications, and future work.

\section{The model and the nonclassicality Verification}
\label{sec:the DRC model and its nonclassicality}

In this section, we first briefly introduce the RCQC model. Then, we show the 1-U-M DRCQC model described by a PAM scenario (details regarding the PAM scenario are provided in the Appendix) and how dimension witnesses for the PAM scenario can be utilized to verify their nonclassical behaviors. 

\subsection{The RCQC model in the prepare-and-measure scenario}
\label{sec:the model}

\begin{figure}
  \includegraphics{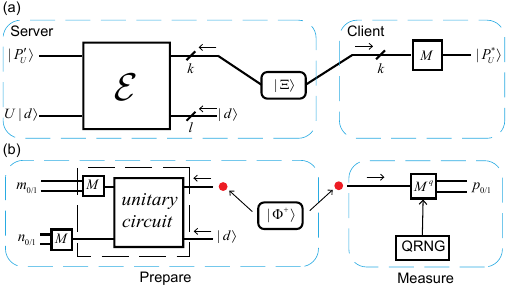}
          \caption{The scheme of remote-controlled quantum computing (RCQC) model. (a) The general RCQC model. The box labeled $\mathcal{E}$ represents a programmable quantum gate array, whose program register and data register consist of $k$ and $l$ qubits respectively. (b) The 1-U-M RCQC model which can be readily described by a PAM scenario. In this model, the programmable gate array $\mathcal{E}$ with $k=l=1$ is realized by a unitary circuit with a single projective measurement made on the program qubit at the output (see the black dashed box). Here, the three symbols in ``1-U-M'' respectively correspond to the condition $k=l=1$, the unitary circuit in the programmable gate, and the measurement acting on the program qubit in the programmable gate. The program qubit is marked by the red dot on the left, and the remote-control qubit by the red dot on the right. To describe this class of models by a PAM scenario, we additionally introduce a measurement on the data qubit at the output stage of the programmable gate array $\mathcal{E}$ and use a QRNG to select one of two measurements ($M^q$ for $q=0,1$) to perform on the remote-control qubit.}
  \label{fig:programmable gate array and its PAM}
\end{figure}

In the RCQC model, quantum entanglement enables the client to achieve remote control over quantum computations delegated to the server~\cite{Qiang_2017,Wang:20}, which are executed via a programmable quantum gate array~\cite{Nielsen_1997} inspired by the classical von Neumann architecture and universal Turing machines~\cite{PRXQuantum.2.030308}. A programmable quantum gate array $\mathcal{E}$ takes input as a data register $\ket{d}$ consisting of $l$ qubits and a program register $\ket{P_U}$ consisting of $k$ qubits that encode the unitary operator to be applied to the data register. Then, it is able to output the desired state $U\ket{d}$ after completing the whole process. Reference~\cite{Wang:20} proposed that for any given programmable quantum gate array, it can be extended to a RCQC model using quantum entanglement (see \cref{fig:programmable gate array and its PAM}(a)). Specifically, the input program qubits $\ket{P_U}$ of the gate array $\mathcal{E}$ are replaced by the first $k$ qubits of a state $\ket{\Xi}=\bigotimes_{i=1}^{k}\ket{\Phi_{i,k+i}^{+}}$ where the state $\ket{\Phi_{x,y}^+}$ is a Bell state $(\ket{0_x0_y}+\ket{1_x1_y})/\sqrt{2}$ shared between qubits $x$ and $y$. Because of projecting the remaining $k$ qubits of $\ket{\Xi}$ onto $\ket{P_U^*}$, the first $k$ qubits collapse to $\ket{P_U}$. Here, $\ket{P_U^*}$ is the complex conjugate state corresponding to $\ket{P_U}$. Thus, we can remotely control the computation performed on the data qubit. Additionally, it is noteworthy that the operation of the gate array $\mathcal{E}$ commutes with the measurement on the last $k$ qubits of $\ket{\Xi}$.

In this work, we consider the 1-U-M RCQC model as shown in \cref{fig:programmable gate array and its PAM}(b). There are two reasons for considering this model. For the first reason, the following two conditions are sufficient to implement the RCQC model with the minimal number of qubits. As the first condition, both the data register and the program register consist of a single qubit, which is the origin of the ``1'' in the 1-U-M notation. Thus, only one pair of qubits in a Bell state $\ket{\Phi^+}$ is required to achieve remote-control of the operators applied to the data qubit. As the second condition, the programmable quantum gate array $\mathcal{E}$ is implemented via a unitary circuit with a single projective measurement made on the program qubit at the output. The ``U'' in 1-U-M refers to the unitary circuit, while the ``M'' refers the projective measurement made on the program qubit. For the second reason, through slight modifications, this model can be readily described by the PAM scenario. For this 1-U-M programmable gate array $\mathcal{E}$, whether the output is the desired state $U\ket{d}$ depends on the measurement result of the program qubit, given that the second qubit of $\ket{\Phi^+}$ is projected onto $\ket{P_U^*}$. 

For convenience, the following conventions are ado\-pted. First, the second qubit of the Bell state $\ket{\Phi^+}$ is referred to as the remote-control qubit. Therefore, three types of qubits appear in this work: the program qubit and the data qubit, which are input to the programmable gate array $\mathcal{E}$, and the remote-control qubit that is tended to be projected onto the state $\ket{P_U^*}$, corresponding to the measurement outcome $p_0$. Second, when the program qubit measurement yields $m_0$, corresponding to the post-measurement state $\ket{P'_U}=\ket{m_0}$, the 1-U-M programmable gate array $\mathcal{E}$ outputs the target state $U\ket{d}$. When the result is $m_1$ ($\ket{P'_U}=\ket{m_1}$), it outputs an alternative state $U'\ket{d}$, where the operator $U'$ depends on the specific implementation of the 1-U-M programmable gate array $\mathcal{E}$. This convention is discussed under the condition that the remote-control qubit is projected onto $\ket{P_U^*}$ (or equivalently, when the program qubit in the state $\ket{P_U}$ is input to the programmable gate array, i.e., $\mathcal{E}[\ket{d}\bigotimes\ket{P_U}]\overset{prob.}{=}(U\ket{d})\bigotimes\ket{m_0}$). Finally, the party executing the complex operations of the gate array $\mathcal{E}$ is treated as the server, while the party performing only simple measurements on the remote-control qubit acts as the client.

The above-mentioned model can be described by the PAM scenario via slight modifications. To describe this model by a PAM scenario, an additional measurement operation is introduced on the data qubit at the output stage of the programmable gate array $\mathcal{E}$ (see \cref{fig:programmable gate array and its PAM}(b)), trying to exhibit its nonclassicality. Analogous to the second convention outlined above, this measurement yields two possible outcomes, denoted as $n_0$ (corresponding to the post-measurement state $\ket{n_0}$) and $n_1$ (corresponding to $\ket{n_1}$). In addition, the following three steps are performed in order. First, the server performs the programmable gate array $\mathcal{E}$ on both the program register and data register. Second, the server measures the data qubit output from the gate array $\mathcal{E}$. Finally, the client makes a projective measurement on the remote-control qubit. Based on quantum mechanics, Steps 2 and 3 are interchangeable because these two measurement operations commute. However, to complete the preparation stage of the PAM scenario before proceeding to its measurement stage, Steps 1 and 2 must be executed prior to Step 3. Moreover, the order in which Step 1 precedes Step 3 satisfies reversed temporal order in quantum information processing. We refer to the 1-U-M RCQC model that obeys this temporal order as the delayed 1-U-M RCQC model.

After completing Steps 1 and 2, the measurement outcomes of the program qubit and data qubit yield four possible combinations---$(m_0,n_0)$, $(m_0,n_1)$, $(m_1,n_0)$, and $(m_1,n_1)$. Here, the first value ($m_i$, $i=0,1$) in parentheses represents the program qubit's measurement result, while the second ($n_i$, $i=0,1$) corresponds to the data qubit's measurement result. The four possible outcomes can be conceptualized as four buttons on the preparation device, which can be represented by the random variable $X\in\{0,1,2,3\}$ with its numerical values used to label these outcomes according to a specific strategy. Each execution of Steps 1 and 2 triggers one of the four buttons, causing the remote-control qubit (sent to the client) to evolve into a specific quantum state. This means that, in each experimental trial, we effectively prepare and transmit a physical system to the client, thereby completing the preparation stage.

Step 3 is performed to complete the final stage of the PAM scenario, namely the measurement stage. When receiving the evolved remote-control qubit, the client shall choose one of two possible projective measurements ($M^q$ for $q=0,1$ in \cref{fig:programmable gate array and its PAM}(b)) to perform on it. Each measurement is determined by the specific unitary operator the experimenter intends to implement on the initial data state $\ket{d}$. Similarly, the two measurement choices ($M^0$ or $M^1$) and the corresponding measurement outcomes ($p_0$ or $p_1$) are represented by the random variables $Y,R\in\{0,1\}$ respectively, with each numerical value serving as a distinct label. It is important to note that a quantum random number generator (QRNG) can substitute for the experimenter's choice selection, as the QRNG equally guarantees measurement choice independence. Through this process, the delayed 1-U-M RCQC model is successfully described by a PAM scenario including 4 preparations and 2 measurements with binary outcomes.

\subsection{Nonclassicality verification}
\label{sec:examine its nonclassicality}

\begin{figure}
    \includegraphics{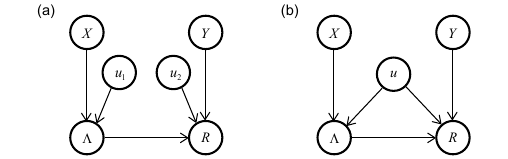}
    \caption{Possible causal structures of the PAM scenario. Classically, the physical system prepared by the preparation device is a random variable $\lambda\in\Lambda$~\cite{PRXQuantum.2.030311}. (a) The first situation that devices are independent and its causal structure represented as DAG. In the DAG of this situation, we only depict the local noise terms (namely $u_1$ and $u_2$) for the random variables $\Lambda$ and $R$. (b) The second situation that devices have preshared correlations and its causal structure represented as DAG.}
    \label{fig:causal structures of the PAM scenario}
\end{figure}

As described below, the PAM scenario has two reasonable situations. Each situation has its corresponding classical causal model. Whether the causal model corresponding to a given situation can be ruled out is determined by checking whether the corresponding dimension witness is violated~\cite{PhysRevLett.112.140407,PhysRevResearch.2.043106}. In both situations, we assume the physical system $\lambda$ sent from the preparation device to the measurement device is \textit{two}-dimensional. Therefore, with this additional dimensional assumption, we verify whether the delayed 1-U-M RCQC model exhibits nonclassical behaviors in a SDI way.

The first situation assumes that preparation and measurement devices are independent \footnote{Reference~\cite{10.1007/s11128-020-02948-3} proposes a method to block communication between devices.}, i.e., $\rho(u_1,u_2)=\rho(u_1)\rho(u_2)$ in \cref{eq:d-classical}. Here, the random variables $u_1$ and $u_2$ represent the internal sources of randomness in the preparation device and the measurement device, respectively, and $\rho(u_1,u_2)$ denotes the joint probability distribution of $u_1$ and $u_2$. Its causal structure is represented by the directed acyclic graph (DAG) shown in \cref{fig:causal structures of the PAM scenario}(a). In the DAG, the nodes represent the random variables, while each directed edge indicates that one variable directly and causally influences the other. In this situation, the random variables $u_1$ and $u_2$ are respectively regarded as local noise terms of the preparation device and measurement device, which are defined as disturbances caused by omitted factors~\cite{J.Pearl2009}. The dimension witness for this situation is nonlinear, making its derivation highly challenging. Remarkably, Ref.~\cite{PhysRevLett.112.140407} provides a dimension witness for the PAM scenario including 4 preparations and 2 measurements with binary outcomes. To construct it, we define a matrix \begin{equation}\label{eq:w}
    \mathbf{W}_2=\begin{pmatrix}
        p(0,0)-p(1,0) & p(2,0)-p(3,0)\\
        p(0,1)-p(1,1) & p(2,1)-p(3,1)
    \end{pmatrix}
\end{equation} from the conditional probabilities $p(r|x,y)$, where $p(x,y)=p(r=0|x,y)$ representing the probability of observing outcome $r=0$ given preparation $x$ and measurement $y$. If an experiment can be explained by this classical causal model, then we always have $\det(\mathbf{W}_2)=0$, regardless of how the preparations and measurement outcomes are relabeled (note that relabeling measurement choices does not affect this result). By contrast, for \textit{two}-dimensional quantum system, the maximum violation of the dimension witness is $\lvert\det(\mathbf{W}_2)\rvert=1$. We assume that the following parts are given: a specific implementation of the programmable gate array $\mathcal{E}$, the two projective measurements on the remote-control qubit, and the initial data qubit $\ket{d}$. If we can find a measurement on the data qubit output from the gate array $\mathcal{E}$, and a labeling strategy for preparations and measurement outcomes such that $\det(\mathbf{W}_2)\neq0$, then we can rule out the classical causal explanation represented by the DAG shown in \cref{fig:causal structures of the PAM scenario}(a) and exhibit its nonclassical behavior in this situation.

In another situation, the preparation and measurement devices have preshared correlations, i.e., $\rho(u_1,u_2)\neq\rho(u_1)\rho(u_2)$ in \cref{eq:d-classical}. In this situation, variables $u_1$ and $u_2$ cannot be treated as local noise terms. For convenience, we represent them collectively as a new composite variable $u=(u_1,u_2)$. The full causal structure is represented by the DAG shown in \cref{fig:causal structures of the PAM scenario}(b). Reference~\cite{PRXQuantum.2.030311} provides a rigorous formulation of the causal assumptions for this situation. First, the preparation and measurement choices must be independent of the variable $u$, i.e., $p(x,y,u)=p(x,y)p(u)$. Second, the future light cone of the preparation device must encompass the measurement device, ensuring that the physical system $\lambda$ sent to the measurement device depends causally only on the input $x$ and the variable $u$. Third, the causal dependence of the outcome $r$ on the input $x$ must be mediated exclusively by the prepared physical system $\lambda$ that is $p(x,r|\lambda,u)=p(x|\lambda,u)p(r|\lambda,u)$. The dimension witness is given by~\cite{PhysRevResearch.2.043106} \begin{equation}\label{eq:I_DW}
\begin{split}
    S=&E_{00}-E_{01}-E_{10}\\
    &+E_{11}-E_{20}-E_{21}+E_{30}+E_{31}\leq4,
\end{split}
\end{equation} where $E_{xy}=p(0|x,y)-p(1|x,y).$ If an experiment admits this classical causal explanation, then the value of $S$ satisfies the bound shown in \cref{eq:I_DW}, regardless of how the preparations are relabeled (note that relabeling measurement choices or outcomes does not affect this bound~\cite{PhysRevResearch.2.043106}). Similarly, we assume that the following parts are given: a specific implementation of the programmable gate array $\mathcal{E}$, the two projective measurements on the remote-control qubit, and the initial data qubit $\ket{d}$. If we can find a measurement on the data qubit output from the gate array $\mathcal{E}$, and a preparation labeling strategy such that $S>4$, then we can rule out the classical causal explanation represented by the DAG shown in \cref{fig:causal structures of the PAM scenario}(b) and exhibit its nonclassical behavior in this situation.

\section{A specific delayed 1-U-M RCQC model and Its Nonclassical Behaviors}
\label{sec:a specific implementation of model}

\begin{figure}
    \includegraphics{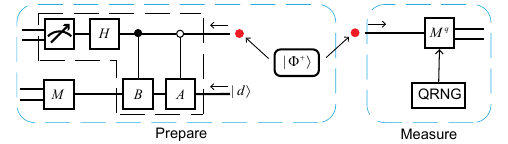}
    \caption{A specific implementation of the delayed 1-U-M RCQC model which we use the methodology outlined in \cref{sec:the model} to describe by a PAM scenario. In this implementation, the programmable gate array $\mathcal{E}$ consists of a controlled-A gate, a controlled-B gate, a Hadamard gate and a computational basis measurement on the program qubit (see the black dashed box).}
    \label{fig:remote-controlled model and its PAM}
\end{figure}

Base on Ref.~\cite{Wang:20}, we design a specific implementation of the delayed 1-U-M RCQC model (see \cref{fig:remote-controlled model and its PAM}) and demonstrate its nonclassical behaviors. For all subsequent operator matrix representations, their input and output bases are exclusively defined in the computational basis. In this scheme, the programmable gate array $\mathcal{E}$ consists of a controlled-A gate, a controlled-B gate, and a Hadamard gate, concluding with a computational basis measurement on the program qubit. For any unitary operator $U$, it can be decomposed as \begin{equation}\label{u_decomposition}
\begin{split}
    U&=e^{i\alpha}\begin{pmatrix}
        e^{-i\frac{\beta+\delta}{2}}\cos(\frac{\gamma}{2}) & -e^{-i\frac{\beta-\delta}{2}}\sin(\frac{\gamma}{2})\\
        e^{i\frac{\beta-\delta}{2}}\sin(\frac{\gamma}{2}) & e^{i\frac{\beta+\delta}{2}}\cos(\frac{\gamma}{2})
    \end{pmatrix}\\
    &=a_0\begin{pmatrix}
        1 & 0\\
        0 & e^{i(\beta+\delta)}
    \end{pmatrix}+a_1\begin{pmatrix}
        0 & -1\\
        e^{i(\beta-\delta)} & 0
    \end{pmatrix}
\end{split}
\end{equation} where $a_0=e^{i(\alpha-\frac{\beta+\delta}{2})}\cos(\frac{\gamma}{2})$, $a_1=e^{i(\alpha-\frac{\beta-\delta}{2})}\sin(\frac{\gamma}{2})$ and $\alpha$, $\beta$, $\delta$, $\gamma$ are real numbers. Then, for the controlled-A gate, the single-qubit gate $A$ is defined with its matrix representation as $A=\left(\begin{smallmatrix}
    1 & 0\\
    0 & e^{i(\beta+\delta)}
\end{smallmatrix}\right)$, while for the controlled-B gate, the single-qubit gate $B$ is defined with matrix representation $B=\left(\begin{smallmatrix}
    0 & -1\\
    e^{i(\beta-\delta)} & 0
\end{smallmatrix}\right)$. The projective measurement of the remote-control qubit is described by $M=\{\ket{P_U^*}\bra{P_U^*},\mathbb{I}-\ket{P_U^*}\bra{P_U^*}\}$, where $\ket{P_U^*}=a_0^*\ket{0}+a_1^*\ket{1}$ is the complex conjugate state corresponding to $\ket{P_U}=a_0\ket{0}+a_1\ket{1}$. When the remote-control qubit is projected onto $\ket{P_U^*}$ and the measurement made on program qubit yields outcome $m_0=0$ (corresponding to the post-measurement state $\ket{m_0}=\ket{0}$), the target state $U\ket{d}$ is obtained in the data register. It is important to note that once the parameters $\beta$ and $\delta$ in the controlled-A and controlled-B gates of the programmable gate array $\mathcal{E}$ are configured, the same parameters $\beta$ and $\delta$ should be consistently applied to the coefficients $a_0$ and $a_1$ in the projective measurement of the remote-control qubit. The term $e^{i\alpha}$ represents a global phase factor and $\alpha$ can therefore be disregarded. For the client, adjusting $\gamma$ in the measurement parameters $a_0$ and $a_1$ enables control over the operator applied to the data register. Hence, this scheme can implement delayed remote control of an infinite (but not complete unless the client modifies the parameters $\beta$ and $\delta$ of the server's controlled unitary gates through communication) set of operators acting on the data qubit, up to a global phase.

Following the methodology outlined in \cref{sec:the model}, this quantum computing model can be described by a PAM scenario (see \cref{fig:remote-controlled model and its PAM}). Specifically, a projective measurement $M=\{\ket{b}\bra{b},\mathbb{I}-\ket{b}\bra{b}\}$ is introduced on the data qubit, where the measurement outcome $n_0$ corresponds to projection onto $\ket{b}$, and $n_1$ corresponds to projection onto the other orthogonal state. Besides, a QRNG is used to randomly select between two measurements $M^q=\{\ket{P_{U_{\gamma_q}}^*}\bra{P_{U_{\gamma_q}}^*},\mathbb{I}-\ket{P_{U_{\gamma_q}}^*}\bra{P_{U_{\gamma_q}}^*}\}$ for the remote-control qubit, where $q=0,1$. Therefore, for a given initial data qubit $\ket{d}=\cos\theta\ket{0}+e^{i\phi}\sin\theta\ket{1}$ and a given labeling strategy, the values of both $\det(\mathbf{W}_2)$ and $S$ can be treated as functions of the independent variables $\beta$, $\delta$, $\gamma_0$, $\gamma_1$, and $\ket{b}$. For a given situation and any given initial data state $\ket{d}$, we expect that it is possible to find all values of the independent variables ($\beta$, $\delta$, $\gamma_0$, $\gamma_1$, and $\ket{b}$) along with a labeling strategy that induce a violation of the corresponding dimension witness---$\det(\mathbf{W}_2)=0$ or $S\leq4$. 

\begin{figure*}
    \includegraphics{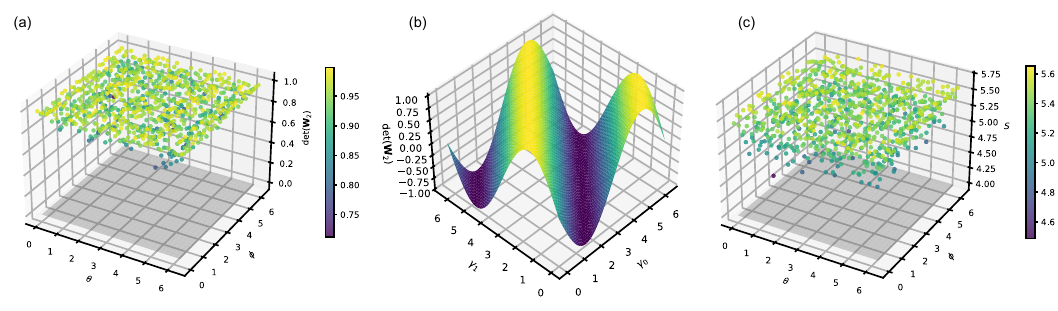}
    \caption{Visualizations of the value of the dimension witnesses. (a) The visualization of the possible maximum values of $\det(\mathbf{W}_2)$ given certain initial data qubit. We use simulated annealing to search for optimal values of independent variables ($\beta$, $\delta$, $\gamma_0$, $\gamma_1$, $\ket{b}$) that maximally violate the dimension witness $\det(\mathbf{W}_2)=0$ for a initial given data qubit $\ket{d}=\cos\theta_i\ket{0}+e^{i\phi_j}\sin\theta_i\ket{1}$ where $\theta_i=\frac{2\pi\times i}{30}$ and $\phi_j=\frac{2\pi\times j}{30}$ ($i,j=0,1,\ldots,29$). All data points are above the gray plane which represents $\det(\mathbf{W}_2)=0$. (b) The visualization of the value $\det(\mathbf{W}_2)$ with respect to the independent variables $\gamma_0$ and $\gamma_1$, while fix $\beta=\pi$, $\delta=0$, $\ket{b}=\cos\frac{\pi}{8}\ket{0}-\sin\frac{\pi}{8}\ket{0}$ and the initial data qubit $\ket{d}=\ket{0}$. (c) The visualization of the possible maximum values of $S$ given certain initial data qubit. We use simulated annealing to search for optimal values of independent variables ($\beta$, $\delta$, $\gamma_0$, $\gamma_1$, $\ket{b}$) that maximally violate the dimension witness $S\leq4$ for a given initial data qubit $\ket{d}=\cos\theta_i\ket{0}+e^{i\phi_j}\sin\theta_i\ket{1}$ where $\theta_i=\frac{2\pi\times i}{30}$ and $\phi_j=\frac{2\pi\times j}{30}$ ($i,j=0,1,\ldots,29$). All data points are above the gray plane that represents $S=4$.}
    \label{fig:dimesion witnesses}
\end{figure*}

Due to the high computational complexity, the simulated annealing algorithm is employed to search for these variable values that can violate the dimension witnesses. Here, we define $\theta_i=\frac{2\pi\times i}{30}$ and $\phi_j=\frac{2\pi\times j}{30}$ where $i,j=0,1,\cdots,29$. Using simulated annealing in the first situation---there are no preshared correlations between the devices and its causal structure as shown in \cref{fig:causal structures of the PAM scenario}(a), we search for optimal values of independent variables ($\beta$, $\delta$, $\gamma_0$, $\gamma_1$, and $\ket{b}$) that maximally violate the dimension witness $\det(\mathbf{W}_2)=0$ for a given initial data qubit $\ket{d}=\cos\theta_i\ket{0}+e^{i\phi_j}\sin\theta_i\ket{1}$ ($i,j=0,1,\ldots,29$). In this situation, the preparation choices $(m_0,n_0)$, $(m_0,n_1)$, $(m_1,n_0)$ and $(m_1,n_1)$ are labeled by the values 0, 1, 2, and 3 of the random variable $X$ respectively, while the measurement outcomes $p_0$ and $p_1$ are labeled by the values 0 and 1 of the random variable $R$ respectively. As described in \cref{sec:the DRC model and its nonclassicality}, the measurement outcome $p_0$ corresponds to the remote-control qubit being projected onto $\ket{P_U^*}$, while $p_1$ corresponds to projection onto another state, and the labeling of measurement choices does not affect the dimension witness. As shown in \cref{fig:dimesion witnesses}(a), all 900 data points (represented by circles) exhibit non-zero values and cluster near 1. 

Besides, for the initial data qubit state $\ket{0}$, the values of independent variables $\beta$ and $\delta$ are additionally fixed to $\pi$ and $0$, along with the measurement $\{\ket{b}\bra{b},\mathbb{I}-\ket{b}\bra{b}\}$ on the data qubit where $\ket{b}=\cos\frac{\pi}{8}\ket{0}-\sin\frac{\pi}{8}\ket{1}$. The labeling strategy is kept unchanged. Then the value of $\det(\mathbf{W}_2)$ can then be treated as a function of the variables $\gamma_0$ and $\gamma_1$. This function can be visualized as the surface plot in \cref{fig:dimesion witnesses}(b), where it is evident that most function values are non-zero. Based on these results, it can be concluded that in most cases under this labeling strategy, for any given initial data qubit $\ket{d}$, there exists a measurement on the data qubit that the dimension witness $\det(\mathbf{W}_2)=0$ will be violated as long as $\gamma_0\neq\gamma_1$, regardless of the values taken by $\beta$ and $\delta$. Thus, the classical causal explanation represented by the DAG shown in \cref{fig:causal structures of the PAM scenario}(a) can be ruled out, thereby exhibiting nonclassical behaviors of the model in this situation.

Similarly, for the second situation---the devices have preshared correlations and its causal structure as shown in \cref{fig:causal structures of the PAM scenario}(b), we use simulated annealing to search for optimal values of independent variables ($\beta$, $\delta$, $\gamma_0$, $\gamma_1$, and $\ket{b}$) that maximally violate the dimension witness $S\leq4$ for a given initial data qubit $\ket{d}=\cos\theta_i\ket{0}+e^{i\phi_j}\sin\theta_i\ket{1}$ ($i,j=0,1,\ldots,29$). To obtain the maximum possible value of $S$, the algorithm is to search for the larger $S$ between the two labeling strategies that assign the tuples $(m_0,n_0)$, $(m_0,n_1)$, $(m_1,n_0)$, $(m_1,n_1)$ to the numerical values 0,1,2,3 and 0,2,1,3 of the random variable $X$, respectively (the labeling of measurement inputs and outputs does not affect the value of the dimension witness). As shown in \cref{fig:dimesion witnesses}(c), all 900 data points strictly exceed the value of 4. Thus, in these cases, the classical causal explanation represented by the DAG shown in \cref{fig:causal structures of the PAM scenario}(b) can be ruled out, thereby exhibiting nonclassical behaviors of the model in this situation. It is reasonable to conjecture that for any arbitrary initial data qubit $\ket{d}$, there exists a measurement on the data qubit and specific parameter values for $\beta$, $\delta$, $\gamma_0$ and $\gamma_1$, such that, under an appropriate labeling strategy, violating the dimension witness $S\leq4$ and exhibiting its nonclassical behaviors.

\section{Conclusion}
\label{sec:conclusion}

In this work, we have shown that the nonclassicality of the delayed 1-U-M RCQC model can be verified in a SDI way. By introducing an additional measurement on the data qubit at the output stage of the programmable gate array, and allowing the client to choose between two measurement choices to measure the remote-control qubit, this model can be described by a PAM scenario. Beyond causal assumptions regarding the PAM scenario, another mild yet crucial assumption is that the physical system transmitted from the preparation device to the measurement device has a dimension of two. We show that classical causal models based on these assumptions can be ruled out by verifying violations of the corresponding dimension witnesses~\cite{PhysRevLett.112.140407,PhysRevResearch.2.043106}. Furthermore, we employ simulated annealing algorithms to search for values of independent variables that violate dimension witnesses ($\det(\mathbf{W}_2)=0$ or $S\leq4$) in a specific model, thereby demonstrating its nonclassicality.

In the main text, we assume that the experimenter knows whether the preparation device and measurement device have or lack preshared correlations. Therefore, to rule out the classical causal model corresponding to a specific situation, it is sufficient to violate the associated dimension witness, thereby demonstrating nonclassical behaviors of the model. Another noteworthy observation is that even without prior knowledge about whether preshared correlations exist between the devices, simultaneous violations of both dimension witnesses $\det(\mathbf{W}_2)=0$ and $S\leq4$ can still conclusively prove the model's nonclassicality.

The delayed 1-U-M RCQC model represents another characteristic and potential of security. In the PAM scenario, the client randomly selects one of two measurements to perform on the remote-control qubit. This approach remains feasible in practical implementations, as one measurement can encode the operator that the client intends to apply to the data qubit, while the other measurement encodes a distinct operator to mislead the server (analogous to decoy states). Furthermore, this strategy can be combined with reversed temporal order in quantum information processing to enhance security and reduce information leakage from the client.

We conjecture that the verification of nonclassicality for the model shown in \cref{fig:programmable gate array and its PAM}(b) can be extended to more general cases, that is, by relaxing the second condition described in \cref{sec:the model}. This only necessitates finding an appropriate approach to describe the model by a PAM scenario and utilizing violations of dimension witnesses to verify its nonclassical behaviors. However, generalizing to higher-dimensional program and data qubits---i.e., to cases where either $k$ or $l$ in \cref{fig:programmable gate array and its PAM}(a) exceeds $1$---faces two main difficulties. The first difficulty is that the model may no longer be straightforwardly described by a PAM scenario in a manner similar to that in \cref{sec:the model}. The second difficulty arises because even if such a description is possible, the number of preparation or measurement choices in this PAM scenario may not match the requirements of pre-derived nonlinear dimension witnesses~\cite{PhysRevLett.112.140407}, and as mentioned in \cref{sec:examine its nonclassicality}, the derivation of nonlinear dimension witnesses is extremely difficult. These topics could be future researches.

\acknowledgments
We thank Carlos de Gois, Rafael Chaves, Gabriela Barreto Lemos, Jacques Pienaar, Romeu Rossi, Yaxuan Wang, and Weixu Shi for fruitful discussions. This work was funded by the National Natural Science Foundation of China (Grant No. 62371459 and No. 62401572) and the Innovation Research Foundation of National University of Defense Technology.

\appendix
\section*{APPENDIX: The Prepare-And-Measure Scenario}
\label{the PAM scenario}
\begin{figure}
    \includegraphics{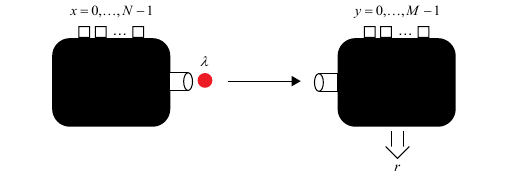}
    \caption{The prepare-and-measure (PAM) scenario represented by black boxes. Upon pressing a button $x\in\{0,1,\ldots,N-1\}$, the preparator sends a physical system $\lambda$ to the measurement device. When receive the physical system $\lambda$ and press the button $y\in\{0,1,\ldots,M-1\}$, it measures different observables based on these inputs and outputs a result $r\in\{0,1,\ldots,L-1\}$.}
    \label{fig:the PAM scenario}
\end{figure}

In the PAM scenario, there are two uncharacterized black boxes as shown in \cref{fig:the PAM scenario}. The first black box contains $N$ buttons. Upon pressing a button $x\in\{0,1,\ldots,N-1\}$, it sends a physical system $\lambda$ to the second black box. The physical system $\lambda$ is empirically unobservable and can therefore be regarded as a hidden variable. The second black box features $M$ buttons. When receiving the physical system and pressing a button $y\in\{0,1,\ldots,M-1\}$, it measures different observables based on these inputs, producing a measurement result $r\in\{0,1,\ldots,L-1\}$. The experiment is described by the conditional probabilities $p(r|x,y)$, representing the probability of observing outcome $r$ given preparation $x$ and measurement $y$. If the experiment admits a $d$-dimensional quantum system representation, then $p(r|x,y)$ can be expressed in the form $p(r|x,y)=tr(\rho_xM_r^y)$ for density matrix $\rho_x$ describing the physical system $\lambda$ and measurement operator $M_r^y$ acting on $\mathbb{C}^d$ $(\Sigma_rM_r^y=\mathbb{I})$. Reference~\cite{PhysRevLett.112.140407} provides the $d$-dimensional classical system representation (i.e., $\lambda=0,1,\ldots,d-1$) for $p(r|x,y)$ as \begin{equation}\label{eq:d-classical}
\begin{split}
   p(r|x,y)=&\int du_1du_2\Big[\rho(u_1,u_2)\\
            &\times\sum_{\lambda=0}^{d-1}p(\lambda|x,u_1)p(r|\lambda,y,u_2)\Big], 
\end{split}
\end{equation} where the random variables $u_1$ and $u_2$ represent the internal sources of randomness in the preparation device and the measurement device, respectively, and $\rho(u_1,u_2)$ denotes the joint probability distribution of $u_1$ and $u_2$.

\def\bibsection{\medskip\begin{center}\rule{0.5\columnwidth}{.8pt}\end{center}\medskip} 
\bibliography{reference}

\end{document}